# Driving Assistance System for Ambulances to Minimise the Vibrations in Patient Cabin

Abdulaziz Aldegheishem, Nabil Alrajeh, Lorena Parra, Oscar Romero and Jaime Lloret

**Abstract:** The ambulance service is the main transport for diseased or injured people which suffers the same acceleration forces as regular vehicles. These accelerations, caused by the movement of the vehicle, impact the performance of tasks executed by sanitary personnel, which can affect patient survival or recovery time. In this paper, we have trained, validated, and tested a system to assess driving in ambulance services. The proposed system is composed of a sensor node which measures the vehicle vibrations using an accelerometer. It also includes a GPS sensor, a battery, a display, and a speaker. When two possible routes reach the same destination point, the system compares the two routes based on previously classified data and calculates an index and a score. Thus, the index balances the possible routes in terms of time to reach the destination and the vibrations suffered in the patient cabin to recommend the route that minimises those vibrations. Three datasets are used to train, validate, and test the system. Based on an Artificial Neural network (ANN), the classification model is trained with tagged data classified as low, medium, and high vibrations, and 97% accuracy is achieved. Then, the obtained model is validated using data from three routes of another region. Finally, the system is tested in two new scenarios with two possible routes to reach the destination. The results indicate that the route with less vibration is preferred when there are low time differences (less than 6%) between the two possible routes. Nonetheless, with the current weighting factors, the shortest route is preferred when time differences between routes are higher than 20%, regardless of the higher vibrations in the shortest route.

## 1. Introduction

In cities, transport is critical, and many efforts are currently being made to improve its efficiency and reach autonomous driving [1]. The transport of patients on ambulances represents a minor percentage of the vehicle classes in urban and interurban mobility. Most proposals for traffic efficiency are tailored for private cars or public transport but do not include emergency vehicles such as ambulances. Nonetheless, the impacts of the traffic, driving patterns, and the vibrations suffered in the patient's cabin are crucial for the patient's well-being and survival.

Most of the proposals for ambulances and mobility are aimed at finding the most suitable location of units to reduce the response time. Thus, the recurrent topics for ambulances and mobility are the forecasting of ambulances demand [2,3], air mobility [4–6], and ambulance allocation [7,8]. Nonetheless, very few proposals are found for the mobility of ambulances after they reach the patient and must bring him to a hospital. An example of these proposals is the doctoral dissertation of J. Miles [9], in which he has developed a decision-support model for transport decisions with respect to a wide variety of patients beyond the traditional time-critical accident and emergency patients. Nonetheless, the approach of this decision-support model is more focused on the patient than the mobility areas.

Many studies demonstrate how vibration in ambulances affects patients. The vast majority of studies are conducted on neonatal transport, which are very sensitive to vibration [10]. Other cases in which the effects of vibration during transport in the patient cabin have been studied have focussed on cardiovascular resuscitation [11,12], ventilation [13], and the intubation process [14]. A completely different study is focused on the effect of ambulance acceleration and deceleration on the heart rate of the patient [15]. Proposals to reduce these variations are concerned with mechanical aspects rather than mobility. Some examples focus on using spinal immobilisation and spinal motion restriction to reduce the head–neck kinematics in the ambulance patient cabin [16]. Other solutions are based on using new hydro–pneumatic suspension [17], which significantly reduces the vibration in the patient compartment of the ambulance.

The aim of this paper is to present a driving assistance system for ambulances based on low-cost sensors, which proposes a recommended route to reach a destination when two or more possible routes can be taken. As far as we are concerned, no similar system has been proposed. The current efforts to improve driving in ambulances are based on simulatorbased training [18] or Self-Organising Maps [19]. The system uses acceleration and GPS data to characterise the routes; generated data is tagged using Artificial Neural Network (ANN) in the cloud. When the ambulance must reach a destination, the different available routes are compared according to the time in which the vehicle remains in areas with different vibration profiles. An index and a score to compare the routes are proposed in the paper. Once the routes are compared, the one with the lowest index value is suggested and displayed on the screen. To train, validate, and test the proposed system, three scenarios and six routes have been driven to gather data. For the training dataset, three repetitions were conducted. The gathered data were tagged into three groups according to the road type. The validation and the test dataset are composed of different routes. All the included routes have both urban and interurban mobility. Different options for data pre-treatment are considered in this study. The main novelty of this proposal is the consideration of vibration in the patient cabin (as the acceleration values) in the proposal of a route, which minimises the discomfort and maximises the well-being of the patient, and facilitates the practices of sanitary personnel. Existing solutions for minimising the vibrations are based on the use of suspension or immobilisation elements [16,17]. In the following, we highlight the contribution and novelty of this paper:

- The use of vibration data in the vehicle is used as input information for a decisionsupport tool for ambulance drivers to select the most appropriate route. So far, no other paper has analysed the effect of vibration in route planning for ambulances;

- The ANN is used to classify the points along a route according to the 3-axis acceleration and velocity. The results of the ANN are used to calculate an index and a score, which also considers the travel time, to compare different routes. As far as we are concerned, no other publication has used this combination of an ANN and an index for the decision;

- The proposed system is tested in real scenarios, including urban and interurban mobility areas with 10 trips and more than 100 min of data registered. Most of the current proposals are based on simulated data rather than gathered data.

The rest of the paper is structured as follows: Section 2 outlines the related work highlighting the differences between existing proposals and the novelty of this paper. The proposed system is detailed in Section 3, including the mobility areas characterisation, the system description, the test bench, the proposed index and score, the ANN, and the methodology. The results, training, validation, and testing of the proposed system are presented and discussed in Section 4. Finally, Section 5 highlights how the results clearly demonstrate the achievement of the aim of this paper.

## 2. Related Work

In this section, the state of the literature related to the proposed system is outlined. The current proposals for vibration monitoring in different types of vehicles are summarised.

On the one hand, we deal with papers aiming to measure the pavement roughness. J. H. Jeong et al., in 2020 [20], proposed using Convolutional Neural Networks (CNNs) for pavement roughness assessment. The data used in this paper was numerically simulated. Their simulations include four types of vehicles and different speeds. They authors also generated road profiles to train, validate, and test their CNN.

Although the data simulations can be very accurate, our system is based on data gathered on roads using an actual vehicle, which is much more precise in its characterisation of the pavement state and its impact on vehicle vibrations. In 2022, Z. Zhang et al. [21] used a smartphone to gather data on three asphalt pavement segments. Using the smartphone data, a genetic algorithm is applied to compute vehicle model parameters and recursive equations for computing the system state variables from which the profile is estimated. Their results estimated the International Roughness Index (IRI) with a relative error below 11%. In 2022, a similar approach was conducted by Y. I. Alatoom and T. I. Al-Suleiman [22]. The authors used a smartphone to gather data and study the impact of pavement age, traffic loading, and traffic volume on the IRI values. Their results suggest that the ANN is a promising tool for predicting the IRI with average errors below 10%. Even though both papers [21,22] aimed to estimate the IRI with accurate results, they were focused on sensing the vibrations to estimate the IRI, not on using the generated values to map and suggest alternative routes intended to minimise those vibrations. More solutions based on the same principle as [21,22] can be found in the review papers [23,24], published in 2022.

On the other hand, several papers measured the vibration sensed in the vehicle in different circumstances to analyse the impact of the pavement or traffic structures on body vibration. Concerning whole-body vibration, two contributions based on simulations are detailed in [24,25]. In 2020, G. Wang et al. [25] studied the comfort of buses based on numerical simulations of road surface roughnes5. Using the same approach, i.e., the simulation of road surface roughness, P. Mucka, in 2021 [26], studied the IRI thresholds based on whole-body vibration in passenger cars. Some simulations were carried out specifically for ambulances [27,28]. These studies employed complex and comprehensive simulations, but this paper aims to classify the roads based on actual data to establish a recommended route. Other papers are found that base their results on actual measurements. M. L. M. Duarte and G. C. de Melo, in 2018 [29], studied the effect of different pavements on body vibration, such as stone paved road samples and asphalt roads. The authors performed the trials with three cars and at five speeds, 20 to 60 km/h. The trials have a duration of 30 to 43 s. Weighing factors from 1 to 1.4 are applied. The authors do not use the 3-axis acceleration independently; they used the most severe axis acceleration and represent the power spectra density results for Ford Fusion as a function of pavement type, vehicle type, and speed. Notwithstanding the efforts and complex test bench of the authors, the paper was focused on urban mobility. For our purpose, the characterisation of both urban and interurban mobility is necessary. In 2022, S. Bruno et al. [30] used low-cost sensors to manage stone pavements with a GIS-based methodology. They gathered data in actual streets with stone pavements in two types of vehicles, a car and a bicycle. The authors established a series of thresholds to predict the comfortability levels for each acceleration. As in the previous case, the provided results are aligned with the proposal of this paper, but they are focused on urban mobility and stone-paved streets. Furthermore, the established thresholds are for passenger comfortability, whereas our aim is to evaluate comfortability for ambulance cabins, which must be much more restrictive. Finally, in 2022, P. Kehoe et al. [31] measured the vibration profiles in a road ambulance using the equivalent acceleration. This is the only study found in which the acceleration in ambulances is measured. The authors measured the acceleration at different points of the patient cabin to find the most suitable position for neonatal infants. Their results showed that the location barely factors when driving on a relatively smooth section of the road. For our purpose, more information is needed for other road surfaces.

The main novelty of the present proposal is that it is based on actual data from both urban and interurban roads gathered during regular driving. The duration of routes is higher than any other presented in the related work, and the wide range of accelerations and speeds are much closer to the actual operation of an ambulance. Moreover, this is the only proposal that recommended a route as part of the driving assistance to ambulance drivers based on the use of an ANN to classify the data, and the calculation of an index and a score for each route. As far as we are concerned, no other paper has dealt with this issue.

## 3. Assumptions of the Proposed Driving Assistance System for Ambulances

In this section, the proposed system is presented. First, we detail the characterisation of mobility areas assumed for this paper. Then, the system proposal in terms of hardware elements is described. In the third subsection, we identify and characterise the different routes included for system evaluation. Next, the AAN and index proposal are shown. Finally, the methodology to test the proposal performance is described.

### 3.1. Mobility Areas Characterisation for Ambulances

Three different mobility areas are identified for the ambulance's movement both in urban and interurban areas attending to the vibrations in the x-, y-, and z-axes. The first mobility area identified for this proposal includes general interurban roads with good pavement. This is the area with lower variation in all axes. On these roads, there is almost no interference with driving.

The second mobility area includes both urban avenues and conventional interurban roads with regular pavement. In this case, more interferences occur in the z-axis; there are few road bumps, a general absence of manhole covers, and we might find plastic transverse rumble strips. Regarding the y-axis, few variations in the direction are found in these types of roads. Finally, for the x-axis, there are more changes in the velocity than in the first mobility area.

The third mobility area, the streets in urban areas, is characterised by the highest vibration in the z-axis due to pedestrian walkways with road bumps and manhole covers. Moreover, there is a lot of variation in the y-axis due to the rotation of the vehicles in both roundabouts and cornering over blocks. The variation of the x-axis is mainly caused by the changes in velocity due to stops, traffic lights, and other elements. Another example of this third zone is the interurban areas with lousy pavement, which causes much perturbation in the z-axis.

The summary of mobility areas included in this proposal, their nomenclature, and their main characteristics can be seen in Table 1. Note that highways are not included since we propose our system for ambulances moving locally. Thus, we do not expect the use of highways. Moreover, if the highway is used, we assume there is no similar route in terms of km and duration.

**Table 1.** Types of mobility areas in this proposal.

| Name of Mobility Area | Acronym | Example |
|---|---|---|
| Mobility area 1 | A1 | Interurban roads with good pavement |
| Mobility area 2 | A2 | Interurban roads with regular pavement; urban avenues |
| Mobility area 3 | A3 | General urban streets; interurban roads with bad pavement |

*3.2. System Description*

The hardware of the proposed system is the same as that proposed in [32]. It is based on two sensors (G-sensor [33] and a GPS sensor [34]) and a node [35]. Moreover, the sensor node contains a battery, a Liquid Crystal Display (LCD), an SD card, a speaker, and a keyboard. The novelty of the proposal is in operation mode. It has two operation modes. In the first, the node gathers data about the acceleration, velocity, and location of the vehicle. The user includes an origin and destination in the second operation mode, and the node proposes the most suitable route. The node operates with the cloud in which data is uploaded, and maps from Google Maps are loaded to consult the roads.

Operation mode 1 is further described in Figure 1. In this operation mode, the user indicates the operation mode using the keyboard. After entering information into operation mode 1, the sensors start gathering data every second. The values of acceleration (in Fax, Fay, and Faz (m/s2)), location (latitude and longitude), and time and velocity in Km/h are automatically stored in the SD card. The LCD displays the values of acceleration during driving. After the driving, the stored data is sent to the cloud and automatically tagged. In addition, the starting and destination points are used to create the new route, which is included in the cloud database over Google Maps public maps. The use of the cloud in vehicular networks is described and analysed in many papers [36,37].

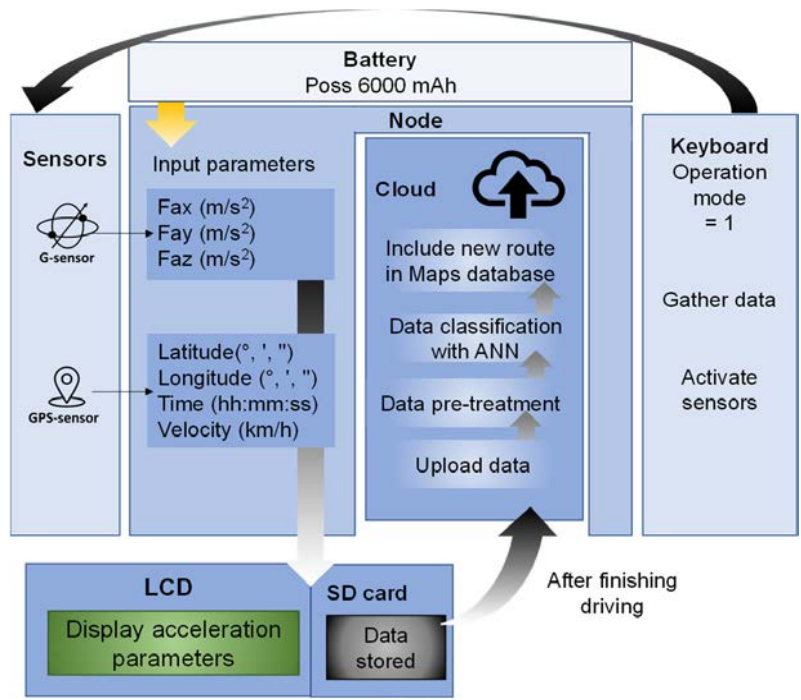

**Figure 1.** Operation mode 1 of the proposed system.

Figure 2 portrays operation mode 2. After selecting the operation mode, the starting and destination points are established in this case. Then, the node connects to the cloud and searches the available routes. After downloading the scores and indexes, the node selects the recommended route, which is displayed on the LCD. Once the route starts, the sensors start to gather data and store it in the SD card. Before the vehicle enters A2 or A3 zones, a speaker alerts the driver to pay attention to the possible elements such as corners, roundabouts, or pedestrian walkways with road bumps.

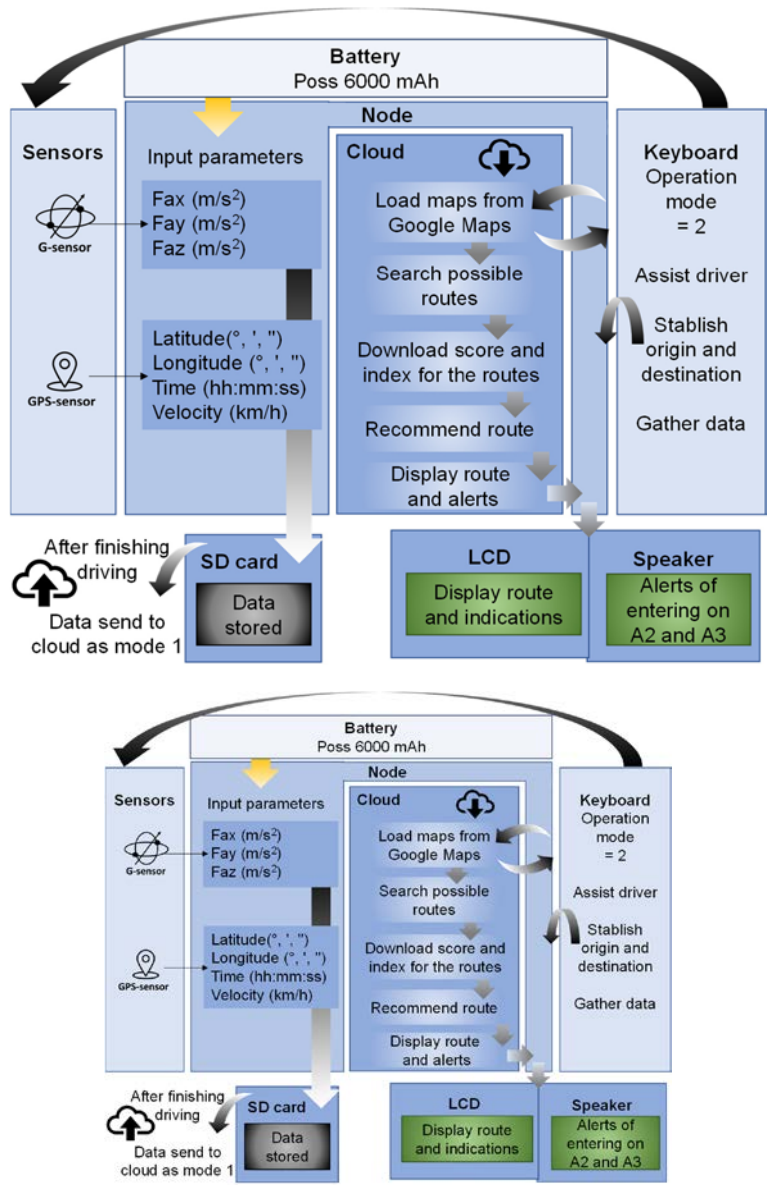

**Figure 2**. Operation mode 2 of the proposed system.

### 3.3. Test Bench

This subsection describes the different routes by which the proposed system has been tested. The proposed system was tested on 11 trips, including different departures, destination points, and routes. The first scenario starts and finishes in urban areas. There are two alternative routes to reach the destination; both routes have a common part, including A1, A2, and A3, represented as a black and grey line in Figure 3. There is a bifurcation in the route where two options appear: both cross different urban and interurban areas. The first option is shorter than the second option in terms of Km and time (see the dark blue line in Figure 3). This option crosses a town longitudinally, and the pavement of interurban areas is lousy. The second option (see the light blue line in Figure 3) is longer than option

1, but it has fewer parts in urban areas, and the pavement of interurban areas is regular. We use portions of the common parts of the route, the grey lines, to train our ANN. The rest of the route, dark blue and light blue, is used for other purposes, detailed in subsequent paragraphs.

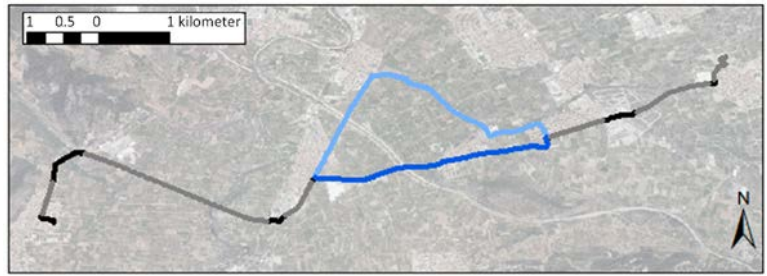

**Figure 3.** The first scenario with two possible routes to reach the same destination**.**

The second scenario comprises three different routes, including different urban and interurban areas in different proportions (see Figure 4). The routes have different lengths and are generated to test different types of pavements and areas of the region in which the system is proposed. We use these routes to verify the ANN model.

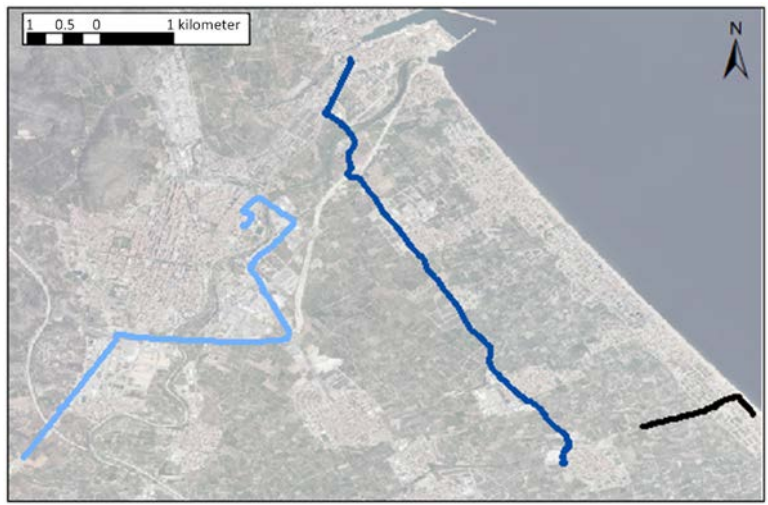

**Figure 4**. The second scenario with different routes to verify the ANN.

The last scenario comprises two alternative routes with the same starting and destination points (see Figure 5). As in the first scenario, both routes share part of the transect. It is a similar case to the first scenario. The light blue line represents a longer transect, but it is composed of interurban roads with excellent pavement, regular pavement, and small urban streets. The second option, the dark blue line, comprises a portion of interurban roads with excellent and regular pavement, urban avenues, and a larger portion of urban streets. The different routes of the third scenario, as well as the different routes of the first scenario, will be used to test the proposed index to decide the most efficient route which minimises the vibrations along the path.

*3.4. Index, Score, and ANN Proposal*

In this subsection, we define the index and the score to include the vibrations suffered in the patient cabin in each route selection. On the one hand, the index is based on weighing the time the ambulance remains on each type of road. The values included in the index are defined to penalise the time remaining in MA3 since the abrupt vibrations which occur in this area might cause a series of problems for patients. This index can be used to compare two routes to decide which one is more suitable in emergencies since the time consumed to travel from departure to destination is considered.

$$Index_{TOTAL\ TIME}\ (s) = A1_{\ TIME}\ (s) + A2_{\ TIME}\ (s) \times 1.5 + A3_{\ TIME}\ (s) \times 2 \qquad (1)$$

where $Index_{TOTAL\ TIME}$ is the total time calculated by the index after weighing the time in the route by the mobility areas, $A1_{\ TIME}$ is the time (in seconds) that the vehicle remains in A1, $A2_{\ TIME}$ is the time (in seconds) that the vehicle remains in A2, $A3_{\ TIME}$ is the time (in seconds) that the vehicle remains in A3, 1.5 and 2 are the weighing coefficients proposed.

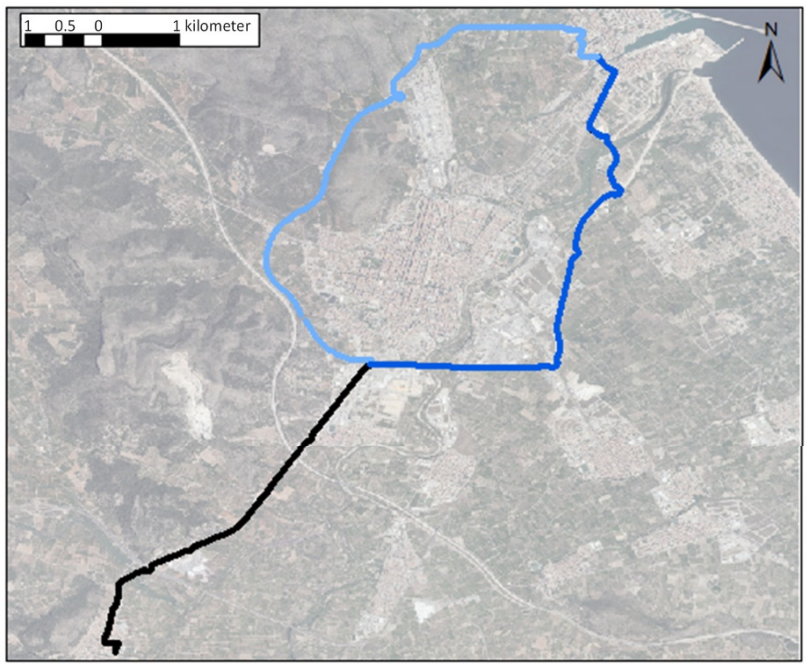

**Figure 5**. The third scenario for system verification.

On the other hand, the score is proposed to compare the effect of vibration in different routes without considering the time. This score is used as a tool by which knowledgeable professionals can improve the system in the future. The score considers the % of route time under each one of the mobility areas and can have values between 1 and 2; the lower the values, the better the route in terms of diminished vibration.

$$\text{Score } (nu) = \frac{1 \times A1_{\text{TIME}}\ (s) + 1.5 \times A2_{\text{TIME}}\ (s) + 2 \times A3_{\text{TIME}}\ (s)}{\text{Total time } (s)} \qquad (2)$$

where Score is the value for the punctuation of the route calculated by the equation according to the time spent in each mobility area during the route.

Concerning the ANN, the input layer is composed of the neurons representing the velocity and acceleration parameters. Since different combinations of parameters will be tested, the number of input neurons might change from three to four. The output layer is composed of three neurons, the mobility areas: A1, A2, and A3. The structure of the ANN is depicted in Figure 6. There are two hidden layers. The established previous probabilities and error cost are the same for all the groups. The sphere of influence was trained using jackknifing; the values of the sphere of influence are presented in the results. The use of ANNs and Convolutional Neural Networks (CNN) are widely used in vehicular networks for preventing accidents [38] and traffic congestion detection [39].

*3.5. Methodology*

In this subsection, the methodology followed to train, validate, and test the system is fully described. The mentioned test bench in Section 3.3 provided a total of 4606 gathered registers. Each data register is composed of the velocity, acceleration in the 3-axis, time, and location.

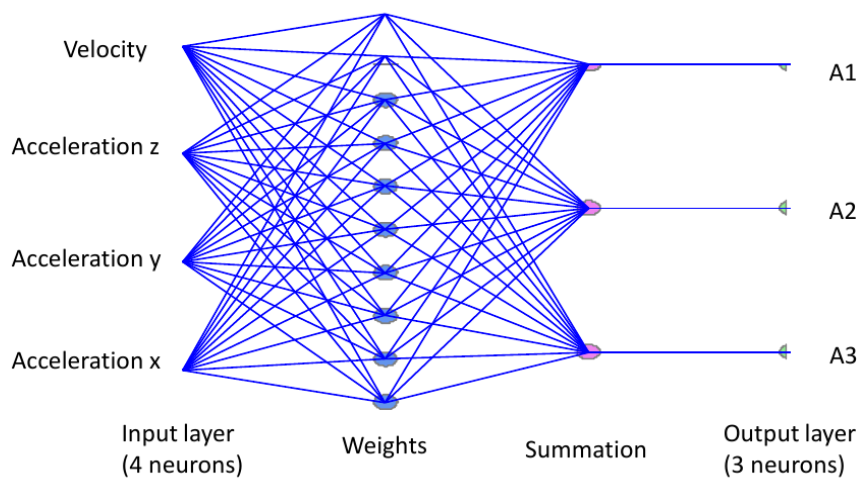

**Figure 6.** Used ANN for data classification.

First, data in the grey colour of Figure 3, with the three repetitions, is used to train the ANN. A total of 1398 registers are used for this purpose. Data is manually tagged according to the pavement, type of area, and presence of elements that produce vibrations. In the ANN training, we tested the effect of data pre-treatment and the necessity of including the 3-axis data. The results of these training tests are illustrated in Section 4.1.

Then, the ANN for data classification is validated with three routes composed of urban and interurban mobility areas. The routes of the second scenario are selected for validating the ANN; they are composed of 2183 registers. The validation is evaluated based on a qualitative approach by checking the assigned class and the real state of the road. Correctly identifying elements such as pedestrian walkways with road bumps, corners, and roundabouts is crucial for the validation success. The results of the validation process are presented in Section 4.2.

Finally, the whole system, including the score and index, is tested in two new scenarios with two available routes for reaching the same destination. The first and third scenarios are used for the final test, comprising a total of 2423 registers. The scenarios include urban and interurban areas and the three possible mobility areas. The results of the complete test of the system are fully described in Section 4.3.

## 4. Results

In this section, we detail the results of testing the proposed system. First, the classification accuracy with different pre-treatment methods before its inclusion in the ANN is presented. After attaining an adequate classification performance of the ANN, the system is validated by classifying different routes. Finally, and to test the operation of the complete system, two routes will be included in the system to allow the ANN to assist driving by offering the best option according to the proposed index.

### 4.1. Data Pre-Treatment and ANN Training Performance

First of all, raw data of the 3-axis and the velocity are tagged as A1, A2, and A3 and included in the ANN. A total of 1736 cases are used for the first ANN training. The performance of this initial ANN model was 77.42% (see the confusion matrix in Table 2). Intending to simplify the system and reduce the possible influence of noise, the vibration of the x and y axes, which are less prejudicial for the patient in the cabin, are different configurations of the ANN. Table 3 summarises the confusion matrix when the x- and z-axis raw data and velocity are used. In this case, the % of cases correctly assigned decreases 1% to 76.04%.

**Table 2.** Confusion matrix using 3-axis raw data and velocity.

| Tagged Mobility Area | Assigned Mobility Area | | | % Correctly Assigned |
|---|---|---|---|---|
| | A1 | A2 | A3 | |
| A1 (779) | 623 (79.97%) | 144 (18.49%) | 12 (1.54%) | |
| A2 (443) | 78 (17.61%) | 322 (72.69%) | 43 (9.71%) | |
| A3 (314) | 6 (1.17%) | 109 (21.21%) | 199 (77.63%) | 77.42% |

Sphere of influence equal to 0.0492188.

**Table 3.** Confusion matrix using x- and z-axis raw data and velocity.

| Tagged Mobility Area | Assigned Mobility Area | | | % Correctly Assigned |
|---|---|---|---|---|
| | A1 | A2 | A3 | |
| A1 (779) | 614 (78.82%) | 162 (20.80%) | 3 (0.39%) | |
| A2 (443) | 80 (18.06%) | 314 (70.88%) | 49 (11.06%) | |
| A3 (314) | 2 (0.39%) | 120 (23.35%) | 192 (76.26%) | 76.04% |

Sphere of influence equal to 0.0507813.

On the other hand, when the data included is from the y- and z-axis raw data and velocity, the performance of the ANN is 77.76%. These results, displayed in Table 4, suppose an improvement compared to using all the data. The improvement confirms that the data of the x-axis imposes noise on the system, and it is better to avoid using it. Considering that we aim to monitor the vibrations in the patient cabin, the most relevant vibrations are due to the z-axis and the y-axis. This principle is aligned with the observed results.

**Table 4.** Confusion matrix using y- and z-axis raw data and velocity.

| Tagged Mobility Area | Assigned Mobility Area | | | % Correctly Assigned |
|---|---|---|---|---|
| | A1 | A2 | A3 | |
| A1 (779) | 623 (79.97%) | 144 (18.49%) | 12 (1.54%) | |
| A2 (443) | 78 (17.61%) | 322 (72.69%) | 43 (9.71%) | |
| A3 (314) | 6 (1.17%) | 109 (21.21%) | 399 (77.63%) | 77.76% |

Sphere of influence equal to 0.0492188.

The high heterogeneity of acceleration data probably causes the relatively low performance of the ANN on the previous paragraph. To improve the performance of the ANN, a pre-treatment technique is performed to integrate the variability of the data. Instead of using the raw data, we propose to use the standard deviation of the raw data. The main reason for doing this pre-treatment is that the vibration is represented by both the acceleration and its variation over time. Thus, the standard deviation of the acceleration of the y and z axes is calculated. Four buffers of time, 5, 9, 15 and 29 s, are selected to calculate the standard deviation. The performance of the ANN increases with the increment of the time buffer. The results of the ANN for buffers of 5, 9, 15, and 29 s can be seen in Tables 5–8, respectively. The number of cases is reduced due to the use of buffers. For the buffer of 5 s (see Table 5) the number of cases is 1645, and the % of correctly assigned cases is 81.95%. The accuracy is maximum for the A3 and minimum for the A2.

**Table 5.** Confusion matrix using y- and z-axis time buffer = 5 s and velocity.

| Tagged Mobility Area | Assigned Mobility Area | | | % Correctly Assigned |
|---|---|---|---|---|
| | A1 | A2 | A3 | |
| A1 (697) | 570 (81.78%) | 123 (17.65%) | 4 (0.57%) | |
| A2 (466) | 83 (17.81%) | 363 (77.90%) | 20 (4.29%) | 81.95% |
| A3 (482) | 3 (0.62%) | 64 (13.28%) | 415 (86.10%) | |

Sphere of influence equal to 0.0492188.

Regarding the buffer of 10 s, the number of cases is 1561 and the % of correctly assigned cases is 88.47% (see Table 6). As in the previous time buffer, the maximum accuracies are reached for A3 and the minimum for A2. For the time buffer of 19 s with 1415 cases (see Table 7) the % of correctly assigned cases is 91.10%; again, the maximum accuracy was obtained for A3. Finally, for the time buffer of 29 s, the number of cases is 1088 and the % of cases correctly classified is 94.34% (see Table 8). Even with the increase in accuracy, we have decided not to use larger time buffers since, in heterogeneous regions, it is common to change from one mobility area to another in short periods.

**Table 6.** Confusion matrix using y- and z-axis time buffer = 9 s and velocity.

| Tagged Mobility Area | Assigned Mobility Area | | | % Correctly Assigned |
|---|---|---|---|---|
| | A1 | A2 | A3 | |
| A1 (661) | 585 (88.50%) | 76 (11.50%) | 0 (0.00%) | |
| A2 (446) | 54 (12.11%) | 372 (83.41%) | 20 (4.48%) | 88.47% |
| A3 (454) | 4 (0.88%) | 26 (5.73%) | 424 (93.93%) | |

Sphere of influence equal to 0.01875.

**Table 7.** Confusion matrix using y- and z-axis time buffer = 15 s and velocity.

| Tagged Mobility Area | Assigned Mobility Area | | | % Correctly Assigned |
|---|---|---|---|---|
| | A1 | A2 | A3 | |
| A1 (597) | 540 (90.45%) | 56 (9.38%) | 1 (0.17%) | |
| A2 (416) | 46 (11.06%) | 359 (86.30%) | 11 (2.64%) | 91.10% |
| A3 (402) | 2 (0.5%) | 10 (2.49%) | 390 (97.01%) | |

Sphere of influence equal to 0.0125.

**Table 8.** Confusion matrix using y- and z-axis time buffer = 29 s and velocity.

| Tagged Mobility Area | Assigned Mobility Area | | | % Correctly Assigned |
|---|---|---|---|---|
| | A1 | A2 | A3 | |
| A1 (455) | 411 (96.92%) | 14 (3.08%) | 0 (0.00%) | |
| A2 (344) | 13 (3.78%) | 330 (95.93%) | 1 (0.29%) | 97.33% |
| A3 (289) | 0 (0.00%) | 1 (0.35%) | 288 (99.65%) | |

Sphere of influence equal to 0.0125.

*4.2. Classified Routes for ANN Validation Performance*

The three routes of the second scenario are used to validate the performance of the ANN proposed in the previous subsection. The validation performance is conducted qualitatively. The comparison of classified points over the routes and the detailed results of urban areas are used to identify the most accurate ANN. We will compare the validation of the ANN using two time buffers, the time buffer of 15 s and the one for 29 s. Those time buffers achieve performances above 90% in cases correctly classified in the previous subsection. The tagged data of the three routes can be seen in Figure 7, when a time buffer of 15 s is used, and in Figure 8, when a time buffer 29 s is used.

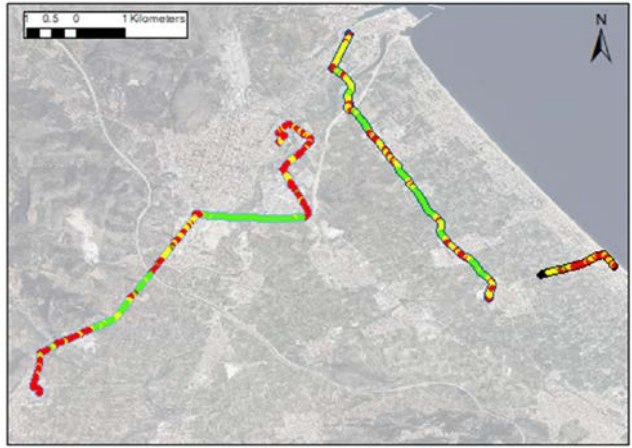

**Figure 7.** Classified points of the three routes of the second scenario using the 15 s buffer.

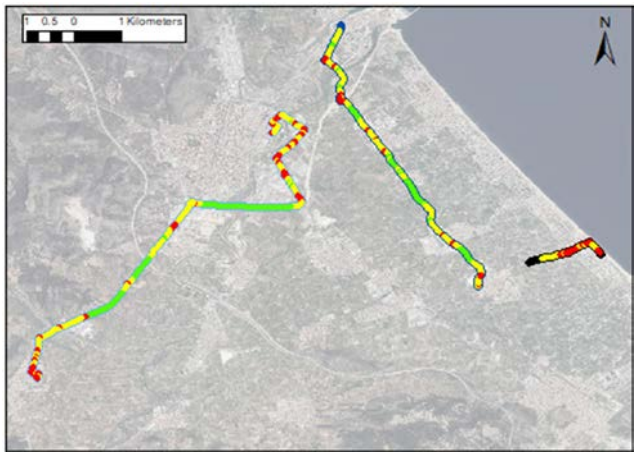

**Figure 8.** Classified points of the three routes of the second scenario using the 29 s buffer.

The first visible differences in classified data with the different ANN models are found for the A3 and A2 classes. In general terms, the A1 is similar to both models. The route indicated in the black line does not include any point classified as A1. According to data classified with the first model, no point is classified as A1. Nonetheless, for the second model (with a time buffer of 15 s), four points are classified as A1. The second route, the one with a dark blue line in Figure 7, comprises A1, A2, and A3 areas. The classification of the points with the first model (ANN with a 15 s of buffer) is 21.0% of A1, 47.5% of A2. and 31.5% of A3. The ANN, with a buffer of 29 s, classified the points as 19.2% of A1, 62.5% of A2, and 12.3% of A3. The main difference between the models is the classification of points from A2 to A3. The classification of A1 points is very similar in both models. Regarding the differences between A2 and A3 points in urban areas, only avenues are included in this route. The main points which can be considered A3 areas are the surroundings of roundabouts and the areas affected by pedestrian walkways with road bumps. A detail of some of the pedestrian walkways with road bumps, marked with black squares, and the classification of A3 points as big red dots, is portrayed in Figures 9 and 10. Figure 10a,b shows the classified points in two urban areas of the second route using the ANN with the 29 s buffer. The detail of the classified points using the ANN with the 15 s buffer is depicted in Figure 9a,b. The results with the ANN and the buffer of 15 s indicate that more regions are classified as A3 when they should be considered as A2 since the effect of pedestrian walkways is extended to adjacent points. This effect is strongly reduced when the buffer of 29 s is used due to the lower weight of the affected point over adjacent points. Thus, the results in the urban areas of this route indicate that using the 29 s buffer has higher accuracy than using the 15 s buffer.

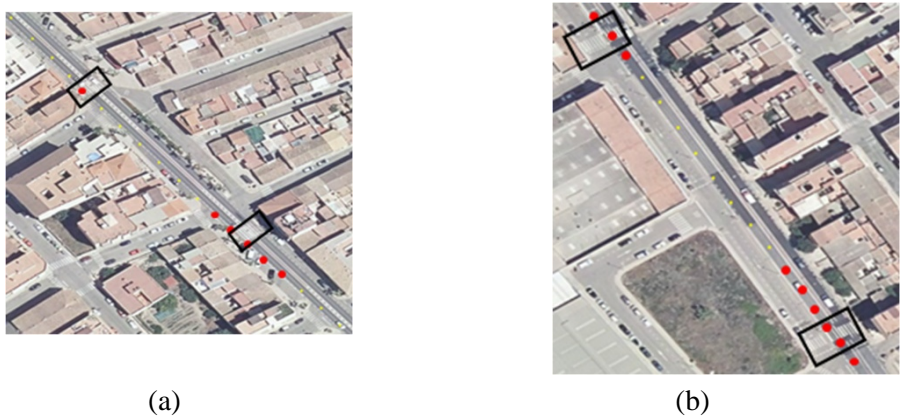

(a)          (b)

**Figure 9**. Detail of classified points of the second route of the second scenario using the ANN with the 15 s buffer, (a) Urban zone of Daimuz, (b) Urban zone of Miramar.

Finally, the third route includes a high percentage of urban areas and A1 roads. Contrary to the second route, which mainly passes over urban avenues, the third route includes urban streets and a typical urban driving style. The classification of the points with the first model (ANN with 15 s of buffer) is 10.7% of A1, 29.9% of A2, and 59.4% of A3. The ANN, with a buffer of 29 s, classified the points as 10.6% of A1, 11.7% of A2, and 77.7% of A3. The classification of A1 points is almost equal for both models. Concerning the differences between the A2 and A3 points, detailed cases are provided in Figures 10 and 11. Figure 11a,b shows the classification of points using the ANN with a buffer of 29 s, while Figure 10a,b represents the results of using the ANN with a 15 s buffer. We identify in red the points classified as A3. It is possible to identify the same effect observed in Figures 8 and 9 in the pedestrian walkways with road bumps, indicated as black squares. In blue squares, we have marked in Figures 11a and 12 the presence of plastic transverse rumble strips, which apparently have no apparently on the classification. Regarding urban driving and turning a street corner, representing A3 areas, the ANN with the 30 s buffer is much more accurate than the second option. When 15 s are used as a buffer, most avenues are classified as A3 instead of as A2. Again, the ANN with a buffer of 29 s offered the most accurate results in the validation.

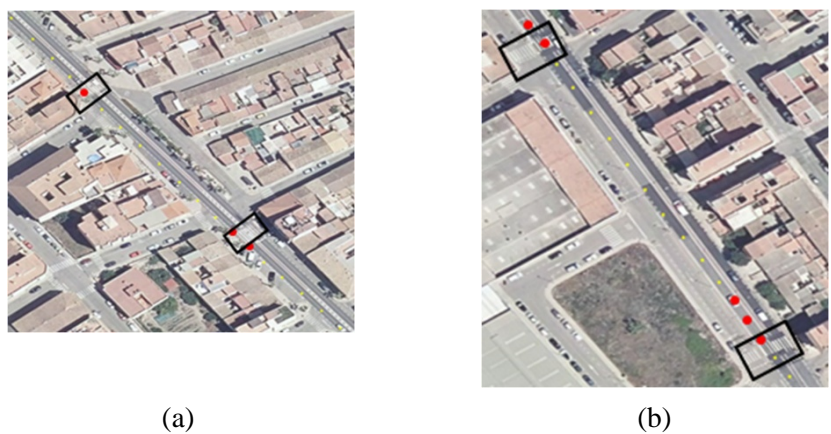

(a)          (b)

**Figure 10.** Detail of classified points of the second route of the second scenario using the ANN with the 29 s buffer, (a) Urban zone of Daimuz, (b) Urban zone of Miramar.

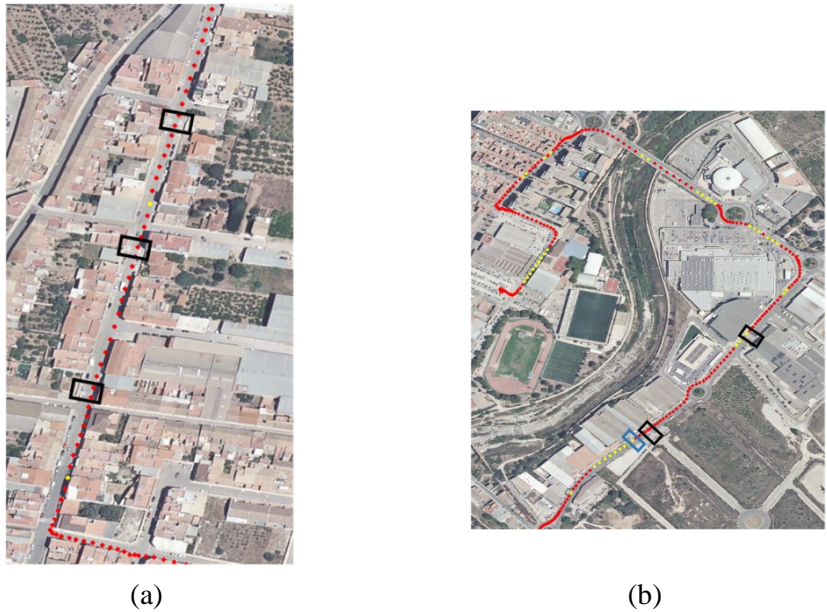

**Figure 11.** Detail of classified points of the third route of the second scenario using the ANN with the 15 s buffer, (a) Urban zone of Gandía, (b) Urban zone of Plama de Gandía.

Considering the pieces of evidence presented in the validation, we have concluded that using 29 s as a time buffer is the most accurate option. Thus, the validation results indicated that the ANN model with the 29 s buffer should be used to calculate the score and index for comparing routes.

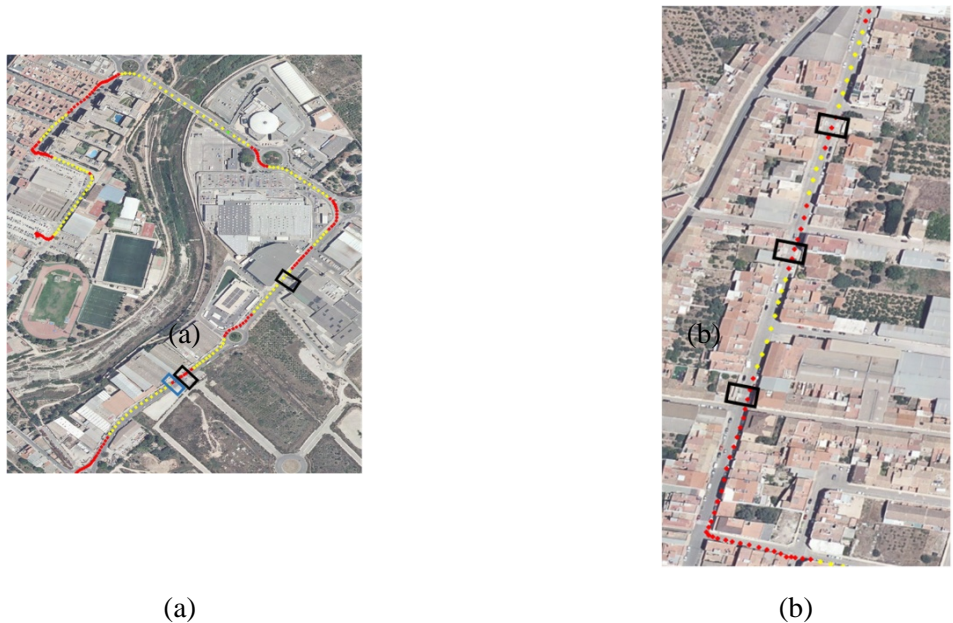

**Figure 12.** Detail of classified points of the third route of the second scenario using the ANN with the 29 s buffer, (a) Urban zone of Gandía, (b) Urban zone of Plama de Gandía.

*4.3. Classified Routes for Index and Score Testing*

The last step is to evaluate the performance of the index and score over two classified routes to recommend the route that minimises the vibration in the patient cabin. The two routes of the first and third scenarios are used for this purpose. First, the data is classified according to the ANN with a buffer time of 29 s. Then, the total travel time and the time over A1, A2, and A3 are estimated to calculate the index and the score. Finally, the index of two available routes is matched to decide which is the recommended route.

The first case in evaluating the index and score is the two possible routes of the first scenario. The first route, dark black in Figure 13, has a total time of 3 min and 40 s (220 s). The vehicle remains 32 s in A1, 95 s in A2, and 135 s in A3. The index calculated for this route is 394 s, and the score 1.8. Regarding the second route, light blue has a total time of 4 min and 22 s (262 s). The time the vehicle remains in A1, A2, and A3 is 21, 50, and 149 s, respectively. The index calculated for the second route is 444.5 s, and the score 1.7. In this case, even though a lower score characterises the first route, it is the preferred route according to the calculated index. The big relative difference in time (42 over 220 s, 20%) and the low difference in the percentage of A3 points (less than 10%) are too high to be compensated with the current weighing factors included in (1). The information is summarised in Table 9.

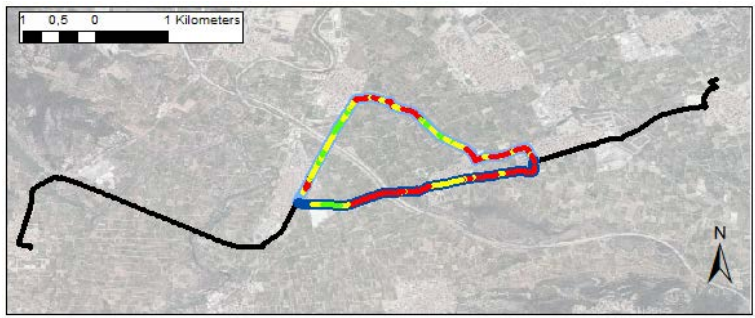

**Figure 13.** Classified routes of the first scenario to test the index and score.

**Table 9.** Results of application of proposed index and scores for the first scenario.

| Route | Total Time (s) | No. A1 | No. A2 | No. A3 | Index (s) | Shorten Route | Preferred Route | Score |
|---|---|---|---|---|---|---|---|---|
| Light blue | 262 | 32 | 95 | 135 | 444.5 | | | 1.70 |
| Dark blue | 220 | 21 | 50 | 149 | 394 | ✓ | ✓ | 1.80 |

The two possible routes of the third scenario are now used as a second case to evaluate the index and score. The first route, dark black in Figure 14, has a total time of 7 min and 45 s (465 s). The vehicle remains 104 s in A1, 164 s in A2, and 197 s in A3. The index calculated for this route is 744 s, and the score 1.49. Regarding the second route, light blue has a total time of 8 min and 13 s (493 s). The time that the vehicle remains in A1, A2, and A3 is 173, 200, and 120 s, respectively. The index calculated for the second route is 444.5 s, and the score 1.43. In this case, although the first route is characterised by a lower total time, the index indicates that the second route is the preferred one. The second route is the one that has the lowest score too. The slight relative difference in time (28 over 465 s, a difference of 6%) and the big difference in the percentage of A3 points over the routes (almost 40%) is high enough to compensate for these 28 s more and make the second route the preferred one, see Table 10.

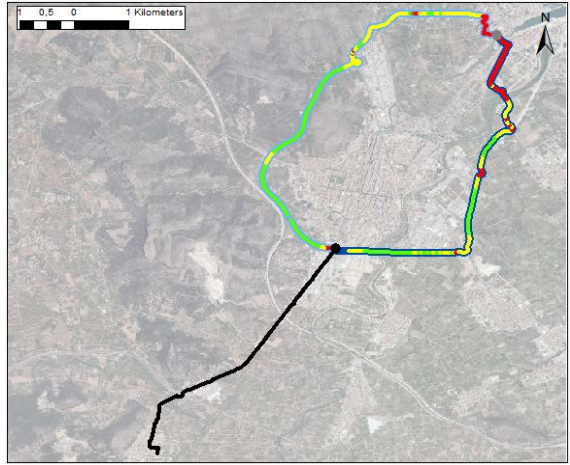

**Figure 14**. Classified routes of the second scenario to test the index and score.

**Table 10.** Results of application of proposed index and scores for the first scenario.

| Route | Total Time (s) | No. A1 | No. A2 | No. A3 | Index (s) | Shorten Route | Preferred Route | Score |
|---|---|---|---|---|---|---|---|---|
| Light blue | 493 | 173 | 200 | 120 | 713 | ✓ | | 1.43 |
| Dark blue | 465 | 104 | 164 | 197 | 744 | | ✓ | 1.49 |

## 5. Conclusions

In this paper, we have proposed a system to assess driving in ambulances to minimise the vibrations in the patient cabin. This proposal aims to reduce the uncomfortability of the patient and facilitate the labour of sanitary personnel if any assistance should be carried out during the travel. The proposed system is based on a node with two operation modes. In the first operation mode, it gathers data from the vibrations during driving, which are later sent to the cloud to tag the points of the route. In the second, the node indicates the destination point and the cloud, using the previous data, Google Maps, and the proposed index and score to suggest a route to the driver. We can affirm that the proposed system achieved our expectations based on the results. We can highlight the following contributions of this proposal:
-  The data of vibration in the x-axis is not useful for data classification with ANN;
- Time buffers are needed for the y-axis and z-axis data to improve the performance of the classification, the 29 s buffer being the best one;
- The ANN to tag the data was trained and validated, achieving a percentage of correctly classified cases up to 97%;
- The proposed ANN, score, and index have been validated over two different scenarios in which two available routes are offered to reach the same destination.
- The following conclusion can be drawn from the obtained results:
- In cases with slight differences in time (less than 6%) over two routes and high differences in pavement typology, the faster route is not always the preferred one;
- The proposed Driving Assistance System for Ambulances can be a promising supportdecision tool;
- The impact of this tool includes the (i) assistance of sanitary managers and drivers to make the most appropriate decision, (ii) a safer service for patients, and (iii) an increase in the comfort of work of medical staff.

In future work, the system will evaluate the effect of different drivers and driver patterns on the classification of data with the ANN and on the results of the index and score. In the same vein, future work will evaluate the impact of different vehicles or the state of the wheels and damper on the gathered data. To improve the performance of the system when additional data is included, the ANN will be compared with other methods such as a kernel extreme learning machine [40], a deep adversarial transfer network [41], and a competitive swarm optimiser [42]. In addition, as pointed out in [43], the problem of black-box models for motion prediction systems will be evaluated. The possibility of including the proposed system in vehicular ad hoc networks [44] will also be considered. In addition, the possible inclusion of information about traffic and weather conditions for longer routes is being studied at the present moment.